\documentclass[aps,preprint,superscriptaddress,groupedaddress,nofootinbib]{revtex4}  

\usepackage{graphicx}  
\usepackage{dcolumn}   
\usepackage{bm}        
\usepackage{amssymb}   
\usepackage{soul}	
\usepackage{color}     
\usepackage{amsmath}

\begin{document}




\title{Low frequency electromagnetic radiation from gravitational waves generated by 
neutron stars}

\author{Preston Jones}
\thanks{Preston.Jones1@erau.edu}
\author{Andri Gretarsson}
\thanks{Andri.Gretarsson@erau.edu}
\affiliation{Embry Riddle Aeronautical University, Prescott, Arizona 86301, USA }
\author{Douglas Singleton}
\thanks{dougs@mail.fresnostate.edu}
\affiliation{California State University Fresno, Fresno, California 93740, USA}

\date{\today}

\begin{abstract}
We investigate the possibility of observing very low frequency (VLF) electromagnetic radiation produced from the vacuum by gravitational waves. We review the calculations leading to the possibility of vacuum conversion of gravitational waves into electromagnetic waves and show how this process evades the well-known prohibition against particle production from gravitational waves. Using Newman-Penrose scalars, we estimate the luminosity of this proposed electromagnetic counterpart radiation coming from gravitational waves produced by neutron star oscillations. The detection of electromagnetic counterpart radiation would provide an indirect way of observing gravitational radiation with future spacecraft missions, especially lunar orbiting probes.  
\end{abstract}

\maketitle


\section{Introduction}

Investigations of electromagnetic radiation associated with gravitational waves usually focus on coincident production at the source where the electromagnetic radiation is of much higher frequency than the gravitational radiation \cite{Connaughton16}.  
It is also possible to generate electromagnetic radiation directly from gravitational waves if the gravitational wave passes through a thin plasma or magnetic field \cite{Macedo83, Marklund00,Lacki10}. The plasma/magnetic field acts as a ``seed" of electromagnetic fields which when acted on by the passing gravitational wave generates additional electromagnetic radiation. The electromagnetic
radiation generated in this way has a frequency that is equal to the frequency of the gravitational wave although the generation of higher harmonic frequencies is also possible \cite{Marklund00}. 

Here we consider a different process: the direct generation of electromagnetic radiation from gravitational waves traveling in the vacuum. This direct, vacuum conversion process does not require a seed plasma or electromagnetic field although one could consider the vacuum fluctuations of the electromagnetic field as the seed field. This vacuum process can be compared to the phenomenon of Hawking radiation where a gravitational background ({\it i.e.} a black hole) can produce quanta of the electromagnetic field ({\it i.e.} photons) from the vacuum. We  find that the electromagnetic radiation from this vacuum production occurs at twice the gravitational wave frequency. 

Unfortunately, electromagnetic radiation generated from a gravitational wave background is expected to have frequencies below the 10~MHz cutoff imposed by the Earth's ionosphere; therefore such low frequency electromagnetic signals are only detectable in space. The Explorer 49 mission~\cite{Douglas85} in 1973 and the previous IMP-6 and RAE-1 missions~(\cite{IMP6RAE1} and references therein)  demonstrated the feasibility of detecting very low frequency (VLF) radiation in a lunar orbit. There is now a resurgence of international interest in missions to lunar orbit as evident by the Indian Chandrayaan-2 and Japanese Selene-2 planned for 2018 and the NASA EM-1 in 2019.  Interest in a new lunar mission for low frequency radio astronomy has been growing (\cite{ULFpaper} and references therein). A new mission to lunar orbit with the capability of receiving VLF in the tens of kHz may allow for detection of the hypothesized electromagnetic counterpart radiation discussed in this paper.

In the next section, we review the relevant theory for vacuum production as presented in previous papers~\cite{Jones15,Jones16}. We include a discussion of how this process evades the usual prohibition against particle production from gravitational waves~\cite{gibbons}. In Sec.~\ref{PenroseScalars}, we estimate the relative electromagnetic and gravitational wave luminosities which allows us to estimate, in Sec.~\ref{Flux}, the luminosity of the electromagnetic counterpart radiation generated by gravitational waves from neutron star $w$-modes. In Sec.~\ref{Detectability}, we discuss the detectability of this electromagnetic radiation and argue that it is not likely to have been detected by current or previous instruments.\footnote{The detection of such radiation would require ongoing VLF monitoring capability in space. Of past missions, the Voyager missions had some capability to detect VLF electromagnetic radiation consistent with production from neutron star oscillations and Voyager did in fact detect signals in the relevant band~\cite{Kurth84}. However, those signals were probably produced by interaction of the solar wind with ions in the outer heliosphere during times of intense solar activity \cite{Kurth03,Webber09}.}

\section{Vacuum production of electromagnetic radiation from a gravitational wave background} \label{TheoryReview}

The Lagrangian density for the electromagnetic field in curved space-time and including source terms is

\begin{equation}
 \mathcal{L}_
{em}  =  - \frac{1}{4}\left( {\partial _\nu  A_\mu   - \partial _\mu  A_\nu  } \right)\left( {\partial ^\nu  A^\mu   - \partial ^\mu  A^\nu  } \right) +J_\mu A^\mu.
\label{emLagrangian}
\end{equation}

\noindent This can be simplified using the Lorenz gauge \cite{Greiner96}, $\partial_\mu A^\mu = 0$, so that for a source free 
Lagrangian one has

\begin{equation}
 \mathcal{L}_{em}  =  - \frac{1}{2}\partial _\mu  A_\nu  \partial ^\mu  A^\nu .
\label{emLagrangianLG}
\end{equation}

\noindent Assuming a plane wave solution for the electromagnetic field, the massless vector field can be expressed in terms of a mode expansion \cite{Greiner96} 

\begin{equation}
A_\mu  \left( {\kappa,\lambda ,x} \right) = \epsilon_{\mu} ^{(\lambda)} \phi ^{(\lambda)} \left( {\kappa ,x} \right),
\label{ModeExpN}
\end{equation}

\noindent where $\epsilon_{\mu} ^{(\lambda)}$ is the polarization four-vector; the label $\lambda = 0, 1 , 2, 3$  gives the possible polarization state; and $\kappa$ represents the field momentum of $A_\mu$. The polarization four-vector satisfies the condition $\epsilon_{\mu} ^{(\lambda)}  \epsilon^{\mu (\lambda ')}  = \eta ^ {\lambda \lambda'}$. Considering only the two, transverse propagating polarizations [ for example $\lambda = 1, 2$ with plane polarization vectors $\epsilon_{\mu} ^{(1)} = \left(0, 1, 0, 0 \right)$ and
$\epsilon_{\mu} ^{(2)} = \left(0, 0, 1, 0 \right)$ ] the Lagrangian density \eqref{emLagrangian} can then be simplified,

\begin{equation}
 \mathcal{L}_{em}  =  - \partial _\mu  \varphi ^* \partial ^\mu  \varphi ,
\label{emLagrangian2}
\end{equation}

\noindent where $\varphi = \frac{1}{\sqrt{2}} (\phi^{(1)} + i \phi ^{(2)})$ is a complex field that is a combination of the two transverse scalar fields $\phi ^{(1,2)}$. The Lagrangian in \eqref{emLagrangian2} is a massless, complex scalar field in Minkowski space-time. We embed this complex scalar field in a general curved background with a metric $g_{\mu \nu}$. The curved spacetime version of the Lagrangian in \eqref{emLagrangian2} then yields the equations of motion for $\varphi$,

\begin{equation}
\frac{1}{{\sqrt{-g}}} \partial_{\mu} \sqrt{-g} g^{\mu \nu} \partial_{\nu}\varphi = 0,
\label{eomvarphi}
\end{equation}

\noindent where $g = det [g_{\mu \nu}]$ is the determinant of the metric. We then take the metric to be a gravitational wave 
background characterized by

\begin{equation}
ds^2 = -dt^2 + dz^2 + a(u)^2 dx^2 + b(u)^2 dy^2.
\label{GWmetric}
\end{equation}

\noindent For simplicity we have assumed only a ``plus" polarization for the gravitational wave. The variable, $u$, in the metric is one of the usual light front coordinates: $u = z-t$ and $v = z+t$. The metric components $a(u)$ and $b(u)$ will be taken as oscillatory functions of $u$ and the determinant of the metric in \eqref{GWmetric} is $\sqrt{-g} = ab$. Using the metric \eqref{GWmetric} in the field equations \eqref{eomvarphi} yields 

\begin{widetext}
\begin{equation}
\left( {b^2 \partial _x^2  + a^2 \partial _y^2  + ab\partial _z \left( {ab} \right)\partial _z  + a^2 b^2 \partial _z^2  - a^2 b^2 \partial _t^2  - ab\partial _t \left( {ab} \right)\partial _t } \right)\varphi  = 0.
\label{emExpand}
\end{equation}
\end{widetext}

\noindent We take the metric functions of the form $a=1+\varepsilon \left(ku \right)$ and $b=1-\varepsilon \left(ku \right)$ where $\varepsilon = h e^{iku}$ and $h$ is some dimensionless amplitude. Near the source of gravitational wave generation we would need to use ``exact solution" for the metric components $a(u), b(u)$ which would require that they satisfy the condition ${\ddot a}/ a + {\ddot b}/ b = 0$ \cite{Schutz} with the dots indicating derivatives with respect to $u$. In this strong field/near zone the use of the plane wave form is questionable. However, in the ``weak field near zone" and ``wave generation region" \cite{Thorne77} where  $h \ll 1$ is satisfied, one can find an approximate solution to order $h^2$ \cite{Jones16} which represents a vacuum state for the scalar field -- the momenta of the field are taken to zero ({\it i.e.} $\kappa \rightarrow 0$) yet one still finds a traveling wave solution for the field $\varphi (u)$ and thus the vector field $A_\mu$. The parameters of this solution depend only on the parameters $h, k$ of the gravitational wave background; (recall that the $\varphi$ field momenta have been set to zero). The solution for the scalar field equation of motion~\eqref{emExpand} with vanishing field momenta $\kappa \rightarrow 0$ is 

\begin{equation}
\varphi \left( {t,z} \right) = A \left( {1 - h^2 e^{2ik\left( {z - t} \right)} } \right)^{ - \frac{1}{2}}  \approx A \left[ 1 + \frac{1}{2}h^2 e^{2ik\left( {z - t} \right)} \right].
\label{OutState}
\end{equation}

\noindent One can determine by direct substitution that $\varphi$ from \eqref{OutState} solves \eqref{emExpand} to order $h^2$. The electromagnetic field solution given in \eqref{OutState} has twice the frequency of the gravitational wave, which implies that the electromagnetic counterpart radiation will have twice the frequency of the gravitational wave which generated it. 
$A$ is a normalization constant which in Ref. \cite{Jones16} was set to $A = \frac{1}{\sqrt{V}}\frac{1}{\sqrt{2 k}}$ in order to calculate the production rate of field quanta. In the next section we use the result in \eqref{OutState} and the Newman-Penrose formalism \cite{Newman61} (as laid out in \cite{Teukolsky73}) to calculate the ratio of vacuum produced electromagnetic flux to gravitational wave flux. In \cite{Teukolsky73} the normalization of $\varphi$ in \eqref{OutState} was taken as $A=1$ which is the normalization we take here.

We now address the apparent conflict between the above result, which in \cite{Jones16} was used to argue that electromagnetic radiation was produced, in vacuum, by a gravitational plane wave, and earlier work \cite{gibbons,garriga,deser} which indicates that particle/field production via gravitational plane waves in vacuum should be prohibited. As mentioned in \cite{gibbons} there are caveats to this prohibition: the fields produced should not be massless and the produced fields should not be moving in the same direction as the incident gravitational wave. The vacuum ``out" solution of \eqref{OutState} violates both these conditions since the field is massless and since it depends only on $u=z-t$, it moves in the same $+z$ direction as the gravitational wave. However simply showing that the present case violates the caveats used to obtain the ``no production" result does not mean there is particle/field production. To this end we turn to the Bogoliubov $\beta$ coefficients which are indicators of whether or not particle/field production occurs. The $\beta$ coefficients for the present case were calculated in \cite{garriga} and found to be
\begin{equation}
\label{b-beta}
\beta_{ij} = \langle u_i ^{out} | u_j ^{in~*} \rangle \propto \delta (k_{-} + l_{-}) ~,
\end{equation}

\noindent where $k_{-} = \frac{\omega - k_z}{2}$ and $l_{-} = \frac{\omega - l_z}{2}$ are the light front momenta of the scalar field before and after\footnote{In \cite{gibbons} and \cite{garriga} a sandwich gravitational wave background was used. The plane wave background of \eqref{GWmetric} was sandwiched between flat space-times. The functions $u_i ^{out}$ and $u_j ^{in}$ are the solutions in the two asymptotic flat regions that are connected to each other through the intermediate plane wave background \eqref{GWmetric}.}; $\omega =\sqrt{{\bf k}^2 + m^2}$ or $\omega =\sqrt{{\bf l}^2 + m^2}$ respectively; and the indices $i,j$ label the momenta of the outgoing and ingoing scalar field quanta. If $m \ne 0$, it is easy to see that $k_{-} + l_{-}$ cannot vanish. If however, as is true in the case considered here, $m=0$ and ${\bf k, l} \to k_z, l_z$ ({\it i.e.} the before and after momenta of the scalar field are purely along the $+z$ direction) then $k_{-} + l_{-}$ vanishes and the Bogoliubov $\beta$ coefficient is nonzero indicating particle/field production.  The conclusion is that the process we describe evades the restriction against particle/field production from a gravitational plane wave by virtue of being massless and having the produced particles/fields traveling in the same direction as the gravitational wave.

\section{Luminosity calculations via Newman-Penrose scalars} \label{PenroseScalars}

The emitted electromagnetic and gravitational wave powers per unit solid angle of emission, are associated with the projection of invariants onto a null tetrad. These projections are identified as the Newman-Penrose scalars \cite{Newman61} for the electromagnetic radiation and the gravitational radiation respectively. The power per unit solid angle of emission for electromagnetic radiation in general is \cite{Teukolsky73,Lehner09}

\begin{equation}
\frac{d E_{em} }{dt d\Omega}  = \mathop {lim}\limits_{r \to \infty } \frac{{r^2 }}{{4\pi }}\left| {\Phi _2 } \right|^2,
\label{emFlux0}
\end{equation}

\noindent where the Newman-Penrose electromagnetic scalar \cite{Newman61,Teukolsky73,Lehner09,Lehner12_86} is, $\Phi _2  = F_{\mu \nu } \bar m^\mu  n^\nu$ and the null tetrads can be identified as \cite{Lehner12_85}

\begin{equation}
\begin{array}{*{20}c}
   {l^\mu   = \frac{1}{{\sqrt 2 }}\left( {1,0,0,1} \right),} & {n^\mu   = \frac{1}{{\sqrt 2 }}\left( {1,0,0, - 1} \right),}  \\
   {m^\mu   = \frac{1}{{\sqrt 2 }}\left( {0,1,i,0} \right),} & {\bar m^\mu   = \frac{1}{{\sqrt 2 }}\left( {0,1, - i,0} \right),}  \\
\end{array}
\label{null1}
\end{equation}

\noindent where

\begin{equation}
 l \cdot n =  - 1,~~m \cdot \bar m = 1 ,~~
l \cdot l = n \cdot n = m \cdot m = \bar m \cdot \bar m = 0.
\label{null2}
\end{equation}

\noindent The electromagnetic tensor can be written as $F_{\mu \nu }  = \partial _\mu  A_\nu  {\kern 1pt}  - \partial _\nu  A_\mu$ where from before, the four-vector potential can be written as $A_\mu   = \epsilon ^{(\lambda)}  _\mu \phi ^{(\lambda )} \left( {t,z} \right)$, again assuming plane polarization $\epsilon ^{(1)} _\mu   = \left(0, 1, 0, 0 \right), \;\epsilon ^{(2)} _\mu   = \left(
0, 0, 1,0 \right).$ The vector field and subsequent electric and magnetic fields in the electromagnetic tensor are found from the derivatives of the scalar field given in Eq. \eqref{OutState}: $\,\partial _t \varphi  =  - ikh^2 e^{2ik\left( {z - t} \right)}$ and $\partial_z \varphi = ikh^2 e^{2ik\left( {z - t} \right)}$. Putting all this together, the Newman-Penrose scalar for outgoing electromagnetic radiation connected with the ``out" state from \eqref{OutState} is \cite{Lehner12_85}

\begin{equation}
\Phi _2  = F_{\mu \nu } \bar m^\mu  n^\nu =
\frac{1}{{\sqrt 2 }}e^{ - i\frac{\pi }{4}} \left( { \partial _z \varphi - \partial _t \varphi} \right) =
i e^{ - i\frac{\pi }{4}} \sqrt{2} k h^2 e^{2ik\left( {z - t} \right)},
\label{emNPscalar0}
\end{equation}

\noindent and the square amplitude is

\begin{equation}
\left| {\Phi _2 } \right|^2  = 2 k^2 h^4.
\label{emNPscalar}
\end{equation}

We now calculate the power per unit solid angle of emission \cite{Teukolsky73,Lehner09} of the outgoing gravitational radiation which is proportional to the Newman-Penrose scalar $\Psi _4$:

\begin{equation}
\frac{d E_{gw} }{dt d\Omega}  = \mathop {lim}\limits_{r \to \infty } \frac{{r^2 }}{{16\pi k^2 }}\left| {\Psi _4 } \right|^2 .
\label{GWluminsity}
\end{equation}

\noindent Using \eqref{GWmetric} the outgoing gravitation plane wave radiation Newman-Penrose scalar in vacuum is \cite{Teukolsky73}

\begin{equation}
\Psi _4  =  - R_{\alpha \beta \gamma \delta } n^\alpha  \bar m^\beta  n^\gamma  \bar m^\delta   = a\partial_{u}^{2} a- b\partial_{u}^{2} b,
\label{Psi}
\end{equation}

\noindent where the partial derivatives are with respect to the light cone coordinate, $u$. Using the weak field limit metric where 
$\varepsilon = he^{iku}$ we find

\begin{equation}
 \Psi _4  =  - 2 hk^2 e^{ik\left( {z - t} \right) } \rightarrow |\Psi _4 | ^2 = 4 h^2 k^4.
 \label{Psi4}
 \end{equation}

\noindent From equations \eqref{emFlux0} and \eqref{GWluminsity} we obtain the ratio of the electromagnetic and gravitational wave powers emitted into some particular direction per unit solid angle as

\begin{equation}
\frac{d E_{em}}{d E_{gw}} = \frac{{\left( {\frac{1}{{4\pi }}\left| {\Phi _2 } \right|^2 } \right)}}{{\left( {\frac{1}{{16\pi k^2 }}\left| {\Psi _4 } \right|^2 } \right)}} \to F_{em}   = 4k^2 \frac{{\left| {\Phi _2 } \right|^2 }}{{\left| {\Psi _4 } \right|^2 }}F_{gw} ~.
\label{LuminosityRatio1A}
\end{equation}

\noindent The first term in \eqref{LuminosityRatio1A} is the ratio of differential energies. These are used to obtain the fluxes ({\it i.e.} power per unit area) $F_{em}$ and $F_{gw}$ of the electromagnetic radiation and gravitational radiation respectively. Finally, substituting the Newman-Penrose scalars from \eqref{emNPscalar} and \eqref{Psi4} we obtain a relationship between these fluxes,

\begin{equation}
F_{em}   = 2 h^2 F_{gw} ~,
\label{LuminosityRatio1}
\end{equation}

\noindent where $h^2$ is the amplitude of the gravitational wave at the point of production. Note that since $F_{gw}\sim h^2$, the overall dependence is $F_{em}\sim h^4$ in the generation zone. 

\section{Flux estimates for neutron star oscillations} \label{Flux}

In this section we will give a rough estimate for the flux, $F_{em}$, of electromagnetic counterpart radiation received at Earth from gravitational waves produced by neutron star oscillations within the Milky Way Galaxy. We will be concerned mainly with neutron star $w$-mode oscillations~\cite{Kokkotas97,Gretarsson11}. Gravitational radiation from $w$-modes is at least an order of magnitude weaker than from $f$-modes but is at a sufficiently high frequency to propagate in the interstellar medium and within our solar system. ($w$-modes span the range $8-16$~kHz while $f$-modes span the range $1-3$ kHz.) 

We use \eqref{LuminosityRatio1} to estimate the electromagnetic flux, $F_{em}$, from a given gravitational wave flux, $F_{gw}$, generated by a neutron star $w$-mode. Since the production of electromagnetic counterpart radiation is determined by the gravitational wave amplitude $h$ at the point of production, we first quote estimates for this quantity at a characteristic distance from the source. We will choose a relatively large characteristic distance so that our estimate for the intensity of the electromagnetic counterpart radiation is conservative. In \cite{Thorne77}, a breakdown is given of different regions around the source (see Fig. 1 of that paper): (i) strong field zone, (ii) weak field near zone, (iii) wave generation zone (this is a combination of strong field zone plus weak field near zone), (iv) local wave zone and (v) distant wave zone. We will take as our characteristic distance, $r=r^{(0)}$, at which to find the characteristic gravitational wave strain, $h=h^{(0)}$, as the edge of the weak field near zone. In terms of the wavelength of the gravitational wave, $r^{(0)} \sim \lambda =\frac{c}{f} \approx 30$ km, where in the last step we have inserted $f \approx 10$ kHz.

Recent searches for the gravitational waves produced by neutron star glitches estimate gravitational wave amplitudes at Earth for $f$-modes on the order of $h\sim10^{-23} $ at $1~\rm{kpc}$  [Eq. (6) \cite{Gretarsson11}] in  and the $w$-mode amplitude is expected to be at least an order of magnitude smaller. Assuming a maximum amplitude for $w$-modes of  $10^{-24}$ at $1~\rm{kpc}$, the strain amplitude as a function of distance, $r$, from the source is

\begin{equation}
h \sim 10^{ - 24} \ \left( {\frac{{1~\rm{kpc}}}{r}} \right).
\label{fmodeh}
\end{equation}

\noindent Inserting $r_{(0)}$ into \eqref{fmodeh}, we get the specific value of the dimensionless amplitude $h^{(0)}$

\begin{equation}
h^{(0)}  \sim 10^{ - 24} \left( {\frac{{ 3 \times 10^{19}~\rm{m}}}{{3 \times 10^4~\rm{m} }}  } \right) = 10^{ - 9} .
\label{fmodeh0}
\end{equation}

As noted above, we intentionally choose a conservatively large distance from the source at which to estimate the characteristic strain amplitude and the value of the amplitude in \eqref{fmodeh0} is indeed considerably smaller than the estimate found in \cite{Jones16} using a different method. [In Ref. \cite{Jones16} the estimate of $h ^{(0)}$ was made by requiring the production rate of electromagnetic counterpart radiation to be ``small" which gave $h ^{(0)} \sim 10^{-5} - 10^{-6}$.]  

The gravitational wave flux near the source can be approximated in terms of $h$ as \cite{Schutz96}

\begin{equation}
F_{gw} ^{(0)} = \frac{{c^3 }}{{16\pi G}}\left| {\dot \varepsilon} \right|^2  = \left( {3 \times 10^{35}~\rm{ \frac{{Ws^2 }}{{m^2 }}}} \right) h^2 f^2 
\sim 3 \times 10^{25}~\rm{\frac{W}{m^2}} ~,
\label{gwFlux_h}
\end{equation}

\noindent where we have $\varepsilon = h e^{iku}$ as in \eqref{emExpand}, and in the last step we have used $f \sim 10$ kHz and $h \sim 10^{-9}$ from \eqref{fmodeh0}. Combining the result from \eqref{gwFlux_h} with Eq. \eqref{LuminosityRatio1} we obtain

\begin{equation}
F_{em} ^{(0)} = 2 \times \left( {10^{ - 9}} \right)^2  \times  3 \times 10^{25}~\rm{\frac{W}{{m^2 }}} \sim 6 \times 10^{ 7}~\rm{\frac{W}{{m^2 }}}.
\label{emFlux}
\end{equation}

\noindent Both $F_{gw} ^{(0)}$ from \eqref{gwFlux_h} and $F_{em} ^{(0)}$ from \eqref{emFlux} are large, consistent with the small characteristic distance from the source $r^{(0)}$ at which most of the production is occurring. Note also that $F_{gw} ^{(0)} \gg F_{em} ^{(0)}$. In other words, the production of counterpart electromagnetic radiation is a very small effect.

If we assume that the neutron star source is at a distance of $1~\rm{kpc}$ from Earth, the typical distance scale used in \eqref{fmodeh}, then the electromagnetic flux seen in the Solar System would be

\begin{equation}
F_{em} = F_{em} ^{(0)} \left( \frac{r}{1~kpc} \right)^2 \sim 6 \times 10^{-23}~\rm{\frac{W}{m^2}} ~.
\label{emFlux2}
\end{equation}

\noindent In the last step we have used $F_{em} ^{(0)}$ from \eqref{emFlux} and $r^{(0)} = 3 \times 10^4$ m for the distance associated with $F_{em} ^{(0)}$. The signal strength in \eqref{emFlux2} is comparable to the strongest pulsar signals, about 6~Jy, assuming a 1~kHz signal bandwidth. We discuss the detectability of such a signal below.

\section{Detectability}
\label{Detectability}

The window of observation for potential conversion of gravitational waves to electromagnetic radiation is greatly restricted by the ionized gases in space \cite{Jester09,Lacki10}, which leads to a range of different plasma cutoff frequencies for different regions. These regions are summarized in Table I. The Earth's ionosphere has a plasma cutoff on the order of 10 MHz so that ground-based observation of extraterrestrial electromagnetic radiation with frequencies less than 10 MHz is not possible. In the interplanetary reaches of the Solar System there is a plasma cutoff frequency due to the solar wind that decreases with distance from the Sun. At the distance of Earth's orbit, this cutoff is in the range of 20-30~kHz \cite{Jester09,Lacki10} so that in interplanetary space near Earth's orbit one cannot detect Galactic signals below 20-30 kHz. At the edge of the Solar System, one reaches the interstellar medium (ISM) which has a plasma cutoff of approximately 2 kHz \cite{Kurth84,Jester09}. Electromagnetic radiation below about 2~kHz cannot propagate through the ISM. There is also attenuation below about 3~MHz due to the Galactic warm ionized medium (WIM) which would prevent the detection of all but the strongest extragalactic or distant Galactic sources below this frequency~\cite{Lacki10}. 

\vskip 0.3cm

\begin{table}[!ht]
\label{Cutoffs}
\centering
\begin{tabular}{|c|c|}
\hline  ~~ Region ~~  & \  ~~ Observable frequency range ~~       
\\   
\hline  On Earth & \ $> \sim10 $ MHz          
\\  
\hline  Interplanetary space  (near Earth's orbit) & \ $>$ $20~\rm{kHz} - 30~\rm{kHz} $  
\\
\hline  Interstellar space (outside the heliosphere) & $ > \sim2 $ kHz 
\\   
\hline
\end{tabular}
\caption{The observable frequency ranges for different regions. These restrictions provide a tight window on where one could potentially observe very low frequency electromagnetic radiation.}
\end{table}

It may be possible to detect a VLF electromagnetic counterpart signal with a flux given by \eqref{emFlux2} via a probe in lunar orbit whose orbit is such that it is periodically occulted from the Sun by the Moon. Such occultation would be required so that VLF noise from of the Sun is blocked. The old Explorer 49 satellite from the 1970s had the ability to collect data below 200 kHz but the lunar orbit was too high to allow complete occultation. At these frequencies, the apparent source size would exceed the size of the lunar disk~\cite{Alexander75}. A new satellite similar to Explorer 49, in a lower orbit, with an improved antenna and receiver system may be able to see a VLF electromagnetic signal of the kind proposed here. The signal flux density would be of a similar order of magnitude as the Galactic background radiation~\cite{Fleishman95} and also about the same as the flux density of white noise generated by a modern low-noise RF amplifier at room temperature.

Ideally, one would like to detect the gravitational waves and the counterpart electromagnetic radiation in coincidence. The electromagnetic counterpart radiation to gravitational waves from $w$-modes is possibly the best candidate for the direct detection of the electromagnetic radiation near Earth. Unfortunately, direct detection of the corresponding $w$-mode gravitational waves themselves is unlikely with the current generation of interferometric gravitational wave detectors. Improved sensitivity to $w$-modes or their harmonics might be present at the free spectral range frequency of the arm cavities (37.5~kHz for Advanced LIGO)~\cite{Giampanis08} but detection will most likely require a major upgrade~\cite{Punturo10}. 

Gravitational waves from $f$-modes are more easily detected with gravitational interferometers due to their lower frequency and higher amplitude\cite{Allen98,Benhar04,Abbott08}. However, the electromagnetic counterpart radiation could not reach Earth due to the plasma cutoffs in Table~\ref{Cutoffs}. This leaves us with the possibility of using a detection of $w$-modes via their hypothesized VLF electromagnetic counterpart as a trigger for coincidental detection of associated $f$-mode gravitational radiation~\cite{Gretarsson11,Weinstein12}. Any process which excites both $f$ and $w$ modes is a potential candidate for coincident detection between these bands. The most common such process is probably a neutron star quake.

An inherently much ``louder" signal in both gravitational and electromagnetic radiation is given by a neutron star-neutron star merger such as the recent observation by the LIGO Collaboration \cite{LIGO-2}. However, that source was at a distance of about 40~Mpc and any VLF signals at Earth would have been reduced in power by 9 orders of magnitude compared to the source we consider above at 1~kpc. Given attenuation by the warm interstellar medium, it seems unlikely that even merger-phase counterpart radiation above $10$~kHz could be detected despite the much larger source amplitude as compared to a star-quake-induced $w$-mode. By virtue of their ubiquity, neutron stars experiencing quakes within a few kiloparsecs of Earth are likely to be a more promising source of VLF counterpart radiation.
 
Similarly, it would not have been possible to observe any VLF from the black hole merger signals seen by LIGO \cite{LIGO}.  Each of the detections were from black hole mergers with gravitational wave frequencies below a few hundred Hz. Even with the frequency doubling between gravitational waves and the electromagnetic counterpart radiation, the counterpart radiation frequencies would still be well below the interstellar cutoff frequency. 

So, while prospects for detection of electromagnetic counterpart radiation from $w$-modes generated in star quakes are promising, there is also the possibility of detecting signals due to gravitational waves of unexpected origin. Current gravitational wave detectors are insensitive to any sources radiating above a few kilohertz. Reception of the electromagnetic counterpart to such sources may be the best way to detect them.

\section{Discussion and conclusions}

The recent detection \cite{LIGO-2} of electromagnetic radiation emitted in conjunction with gravitational waves from a neutron star-neutron star merger has led to excitement at the prospect of ``multimessenger" astrophysics where one gets information from different types of radiation -- gravitational and electromagnetic. In this article we have proposed a potentially new type of joint gravitational wave/electromagnetic wave signal based on the vacuum production of electromagnetic waves from a gravitational wave background. This type of joint signal is similar to the seeded production of electromagnetic waves where a gravitational wave creates electromagnetic radiation by passing through a region containing a plasma/magnetic field. In the Appendix we review some estimates for the strength of the electromagnetic waves from seeded production and compare this with our proposed vacuum production. The general conclusion is that seeded production would give a stronger signal of VLF electromagnetic radiation but the systems that could produce a substantial electromagnetic signal via seeded production are much less common than systems that could produce vacuum production. Prohibitions on particle production by gravitational waves \cite{gibbons} and the subsequent attenuation of gravitational plane waves \cite{deser} in vacuum do not apply to the production of electromagnetic radiation from the gravitational waves described here. The production of massless particles/fields from gravitational radiation is consistent with kinematic restrictions \cite{Modanese95} as well as quantum effects restrictions \cite{gibbons}.   

Coincident detection of gravitational wave $f$-modes and vacuum produced VLF electromagnetic radiation coming from $w$-modes is possible. Sensitivities for detection of the gravitational wave $f$-modes are near the limit of current detectors and require only small improvements for future detection. Detection of the VLF electromagnetic radiation produced by $w$-modes depends on the instrumentation and orbits of future lunar orbiters. Instrumentation similar to Explorer 49 and possibly lower orbits for improved occultation could allow coincident detection of gravitational waves from $f$-modes and vacuum production electromagnetic radiation from $w$-modes. 

Finally, we note that rough estimates of electromagnetic counterpart radiation from gravitational waves emitted during core-collapse supernovae should be much higher than those presented here due to neutron star quakes or glitches.  At the appropriate distance from a core-collapse supernova ({\it i.e.} in the near field weak zone), the gravitational wave strain from core collapse~\cite{Kokkotas97} is almost 5 orders of magnitude greater than the gravitational wave strain from glitch-induced $w$-modes. Since we are still in the weak field regime, $h\ll 1$, vacuum production of electromagnetic radiation from gravitational waves goes like $h^4$, leading to the counterpart electromagnetic flux at Earth from a supernova at 50~kpc about 15 orders higher than from $w$-modes at 1~kpc, or about $10^{16}$~Jy ($0.1~{\rm nW m^{-2}Hz^{-1}}$) assuming similar bandwidths. Indeed, given such large flux one might also expect to see extragalactic (local cluster) supernovae with GJy-scale flux at Earth. However, as mentioned earlier, the Galactic WIM attenuates extragalactic and distant Galactic signals at frequencies below about 3~MHz. Yet, given the fluxes involved, it seems possible that electromagnetic counterpart radiation from a Galactic supernova at 50~kpc would be visible, despite the attenuation. Also, any processes that enable upconversion of these low frequency photons to higher frequencies that can travel unhindered are a potentially interesting avenue of study. All electromagnetic counterpart radiation from supernovae would be expected to be ``prompt'' -- it would reach Earth on a similar time frame as the gravitational wave emission itself. There is no time delay in the creation of the VLF electromagnetic radiation since it occurs in vacuum and once created there should be no delay assuming the
electromagnetic radiation is above the relevant cutoffs of the intervening space.Core-collapse supernovae in our Galaxy are rare ($\sim1$ per century) but the possibility of detecting all the radiation types emitted [gravitational waves, prompt electromagnetic counterpart radiation (if present), neutrinos, and the traditional light curve] is an exciting prospect.

\appendix*
\section{Seeded production versus vacuum production}
We now compare the postulated vacuum production of electromagnetic counterpart radiation to the production from a preexisting seed magnetic field. While we have not found any calculation in the literature of ``seeded" production of electromagnetic counterpart radiation from an isolated neutron star undergoing a star quake, Marklund {\it et al.} \cite{Marklund00} considered the closely related case of a binary neutron star merger in the presence of a strong magnetic field. The neutron stars were each taken to have one solar mass and were separated by 20 times the  Schwarzschild radius of the Sun ($\approx 60$ km). For such a binary system the frequency of the gravitational waves emitted was $\sim 10^2$ Hz. It was then assumed that the generation of electromagnetic counterpart radiation from the emitted gravitational wave started at a distance of about 60 times the Schwarzschild radius ($\approx 120$ km). This was because at closer distances the approximations used by Marklund {\it et al.} did not apply. With this setup it was found that the maximum electric field was proportional to the product of the gravitational wave amplitude and the surface magnetic field of the neutron stars 
\begin{equation}
E_{max}  \propto  h_0 ~ B_{surface} \rm{\frac{V}{m }} ,
\label{Emax}
\end{equation}
where $h_0$ is the gravitational wave amplitude at the distance of 60 Schwarzschild radii and $B_{surface}$ is the magnetic field strength at the surface of the neutron stars. Taking $h_0 \sim 0.001$, $B_{surface} \sim 10^8$ T and using \eqref{Emax} it was found that at a distance of 120 times the Schwarzschild radius ($\approx 360$ km) the maximum electric field and the associated flux of electromagnetic counterpart radiation for this example system were
\begin{equation}
E_{\max }  = 50~\rm{\frac{{MV}}{m}} ~~~ \rightarrow ~~~ S = \frac{1}{{2c\mu _0 }}E^2 _{\max }  \sim 10^{12}~\rm{\frac{W}{{m^2 }}} ~.
\label{Emax-S}
\end{equation}
Comparing the flux from the seeded production example given in \eqref{Emax-S} with the flux from the vacuum production example given in \eqref{emFlux} one finds that seeded production in this case is $10 ^4 - 10^5$ larger than vacuum production. This makes sense since one would expect that the flux would be larger when one has a preexisting field to work with. However by lowering the value of $B_{surface}$ the two fluxes of electromagnetic radiation from seeded production versus vacuum production would move closer together in magnitude. 

Despite the lower flux of the hypothesized vacuum production of electromagnetic radiation from gravitational waves we argued in Sec.~\ref{Detectability} that such a vacuum flux would nevertheless be detectable if the source is close enough. The scenario of production of electromagnetic waves from the vacuum by gravitational waves associated with neutron star quakes has the advantage that it would be much more common as compared to seeded production from a binary neutron star system with a strong magnetic field.


\begin{thebibliography}{99}

\bibitem{Connaughton16} V. Connaughton et al, Astrophys. J. {\bf 826} L6, (2016).

\bibitem{Macedo83} P.G. Macedo and A. H. Nelson, Phys. Rev. D {\bf 28}, 2382 (1983).

\bibitem{Marklund00} M. Marklund, G. Brodin, and P. K. S. Dunsby, Astrophys. J. {\bf 536}, 875 (2000), arXiv:astro-ph/9907350v2.

\bibitem{Lacki10} B. C. Lacki, Mon. Not. R. Astron. Soc. {\bf 406}, 863 (2010).

\bibitem{Douglas85} J. N. Douglas and H. J. Smith, Lunar Bases and Space Activities of the 21st Century, edited by W.W. Mendell
(National Aeronautics and Space Administration, Lunar and Planetary Institute, Houston, 1985), pp. 301–306.

\bibitem{IMP6RAE1} R.G. Stone, Space Sci. Rev., {\bf 14}, 534 (1973).

\bibitem{ULFpaper} R.T. Rajan, A-J. Boonstra, M. Bentum, M. Klein-Wolt, F. Belien, M.Arts, N. Saks, A-J. van der Veen, Experimental Astronomy, {\bf 41}, 271 (2016).

\bibitem{Jones15} P. Jones and D. Singleton, Int. J. Mod. Phys. D {\bf 24}, 1544017 (2015).

\bibitem{Jones16} P. Jones, P. McDougall, and D. Singleton, Phys. Rev. D {\bf 95}, 065010 (2017).

\bibitem{gibbons} G.W. Gibbons, Commun. Math. Phys. {\bf 45}, 191 (1975).

\bibitem{Kurth84} W. S. Kurth, D. A. Gurnett, F. L. Scarf, and R. L. Poynter, Nature (London) {\bf 312}, 27 (1984).

\bibitem{Kurth03} D. A. Gurnett, W. S. Kurth, and E. C. Stone, Geophysical Research Letters {\bf 30} 23, 2009 (2003).

\bibitem{Webber09} W. R. Webber and D. S. Intriligator, arxiv:0906:2746 (2009).

\bibitem{Greiner96} W. Greiner and J. Reinhardt, {\it Field quantization}, (Springer-Verlag, Berlin Heidleberg 1996) 154.

\bibitem{Schutz} B. F. Schutz, {\it A first course in general relativity, 2nd edition}, 
(Cambridge University Press, Cambridge 2009) 210-212.

\bibitem{Thorne77} K. S. Thorne, in The Generation of Gravitational Waves: A Review of Computational Tecniques, edited by V. De
Sabbata and J. Weber, Topics in Theoretical and Experimental Gravitation Physics. NATO Advanced Study Institutes Series (Series B: Physics), Vol. 27 (Springer, Boston, MA, 1977).

\bibitem{Newman61} E. Newman and R. Penrose, J. Math. Phys. {\bf 3}, 566 (1962).

\bibitem{Teukolsky73} S. A. Teukolsky, The Astrophys. J. {\bf 185}, 635 (1973).

\bibitem{garriga} J. Garriga and E. Verdaguer, Phys. Rev. D {\bf 43}, 391 (1991).

\bibitem{deser} S. Deser, J. Phys. A {\bf 8}, 1972 (1975).

\bibitem{Lehner09} P. M{\"o}sta, C. Palenzuela, L. Rezzello, L. Lehner, S. Yoshida and D. Pollney, Phys. Rev. D {\bf 81}, 064017 (2010).

\bibitem{Lehner12_86} L. Lehner, C. Palenzuela, S. L. Liebling, C. Thompson, and C. Hanna, Phys. Rev. D {\bf 86}, 104035 (2012).

\bibitem{Lehner12_85} M. Zilh{\~ a}o, V. Cardoso, C. Herdeiro, L. Lehner, and U. Sperhake, Phys. Rev. D {\bf 85}, 124062 (2012).

\bibitem{Kokkotas97} K. D. Kokkotas, ``Stellar pulsations and gravitational waves", Mathematics of gravitation, part 2, gravitational wave detection, Banach center publications, vol. 41, Institute of Mathematics, Polish Academy of Sciences, 31-41 (Warsaw, Poland 1997).

\bibitem{Gretarsson11} J. Abadie {\it et al.}, Phys. Rev. D {\bf 83}, 042001 (2011).

\bibitem{Schutz96} B. F. Schutz, Class. Quantum Grav.,  {\bf 13}, A219 (1996).

\bibitem{Jester09} S. Jester and H. Falcke, New Astron. Rev., {\bf 53}, 1 (2009).

\bibitem{Alexander75} J. K. Alexander, M. L. Kaiser, J. C. Novaco, F. R. Grena, and R.R. Weber, Astron. \& Astrophys., {\bf 40}, 365 (1975).

\bibitem{Fleishman95}  G. D. Fleishman \& Y. V. Tokarev, Astron. \& Astrophys., {\bf 293}, 565 (1995).

\bibitem{Giampanis08} S. Giampanis, {\it Search for a high frequency stochastic background of gravitational waves}, PhD. Thesis, University of Rochester, ProQuest Dissertations Publishing, 2008.

\bibitem{Punturo10} M. Punturo {\it et al.}, Class. Quant. Grav, {\bf 27}, 084007 (2010).

\bibitem{Allen98} G. Allen, N. Andersson, K. D. Kokkotas, and B. F. Schutz, Phys. Rev. D {\bf 58}, 124012 (1998).

\bibitem{Benhar04} O. Benhar,  V. Ferrariand L. Gualtieri, Phys. Rev. D {\bf 70}, 124015 (2004).

\bibitem{Abbott08} B.P.Abbott {\it et al.}, Phys. Rev. Letts. {\bf 101}, 211102 (2008).

\bibitem{Weinstein12} A. J. Weinstein, Class. Quantum Grav. {\bf 29}, 124012 (2012).

\bibitem{LIGO-2} B. P. Abbott {\it et al.} (LIGO Scientific Collaboration and Virgo Collaboration)
Phys. Rev. Lett. {\bf 119}, 161101 (2017).

\bibitem{LIGO} B. P. Abbott {\it et al.} (LIGO Scientific Collaboration and Virgo Collaboration) Phys. Rev. Lett. 116, 061102 (2016); {\it ibid.} Phys. Rev. Lett. {\bf 116}, 241103 (2016); {\it ibid.} Phys. Rev. Lett. {\bf 118}, 221101 (2017).

\bibitem{Modanese95} G. Modanese, Phys. Lett. B {\bf 348}, 51 (1995).

\end{thebibliography}
\end{document}